# Acoustic Dirac degeneracy and topological phase transitions realized by rotating scatterers

Xinhua Wen,[1] Chunyin Qiu,[1*] Jiuyang Lu,[1] Hailong He,[1] Manzhu Ke,[1] and Zhengyou Liu[1,2*]

[1]Key Laboratory of Artificial Micro- and Nano-structures of Ministry of Education and School of Physics and Technology, Wuhan University, Wuhan 430072, China

[2]Institute for Advanced Studies, Wuhan University, Wuhan 430072, China

**Abstract:** The artificial crystals for classical waves provide a good platform to explore the topological physics proposed originally in condensed matter systems. In this paper, acoustic Dirac degeneracy is realized by simply rotating the scatterers in sonic crystals, where the degeneracy is induced accidentally by modulating the scattering strength among the scatterers during the rotation process. This gives a flexible way to create topological phase transition in acoustic systems. Edge states are further observed along the interface separating two topologically distinct gapped sonic crystals.

*Corresponding authors. cyqiu@whu.edu.cn ; zyliu@whu.edu.cn

## I. Introduction

Dirac point, a nodal point degeneracy connected with linear-dispersion in momentum space, has attracted broad attention in condensed matter systems after the realization of the monolayer graphene [1-3], because of many intriguing electronic transport properties (e.g., ultrahigh mobility). The Dirac degeneracy can be generated either deterministically or accidentally [4]. In general, the deterministic degeneracy is a consequence of the high crystal symmetry, and the accidental degeneracy is generated by fine-tuning parameters (e.g., external fields). Interestingly, the system with Dirac degeneracy is a good starting point for investigating topological physics, since it corresponds to the critical point of topological phase transition.

Recently, the classical analog of the unique physics associated with Dirac points has been emerging as an exciting field. For artificial crystals that work on classical waves, both the spatial symmetry of the crystal and the scattering (coupling) strength among the scatterers can be tailored flexibly. Therefore, the classical system, e.g., photonic crystal or sonic crystal (SC), forms a good platform for exploring the Dirac cones and associated topological physics proposed originally in condensed matter systems. In addition to the deterministic twofold Dirac degeneracies [5-7] appearing in the corner points of the hexagonal Brillouin zone (BZ), accidental triple [8-11] and quadruple [12-15] Dirac degeneracies, which are realized by fine-tuning parameters, have also been observed in the hexagonal BZ center. Those systems, carrying zero refraction indices [8-10,16], have been exploited for wavefront shaping and cloaking. By opening the accidental Dirac degeneracy, topologically distinct insulating phases have been observed and utilized to construct topologically protected edge states [15,17,18].

Note that in previous studies, isotropic scatterers are considered and the accidental degeneracy is realized by fine-tuning the filling ratio or material parameters of the scatterers. In this paper, we consider the SCs made of anisotropic scatterers and propose a rotating-scatterer mechanism to realize the accidental Dirac degeneracy. The physics is clear: by simply rotating the scatterers, the scattering strength among the scatterers can be tuned gradually, which enables the closure of bandgap at some critical orientations. Specifically, we consider the SC made of a hexagonal array of fan-like scatterers with three-fold rotation symmetry. During the process of rotating scatterer, the bandgap may close at fixed momenta, i.e., the corners of the hexagonal BZ, where the band repulsion does occur since the degenerate states carry different eigenvalues of the rotation operator. Interestingly, the SC undergoes acoustic valley Hall phase transition during the rotation process. Furthermore, we have observed acoustic edge states traveling along the interfaces separating topologically different acoustic valley Hall insulators. Extension has been made to the SC consisting of snowflake-like scatterers with six-fold rotation symmetry. Accidental quadruple Dirac degeneracy has been realized in the hexagonal BZ center, together with a demonstration of the acoustic quantum spin Hall phase transition and related nontrivial edge states. Throughout this paper, a commercial finite-element software (COMSOL Multiphysics) is applied to perform full-wave simulations, where the solid scatterers placed in air background are modeled as acoustically rigid.

**II. Accidental double Dirac degeneracy and acoustic valley Hall phase transition**

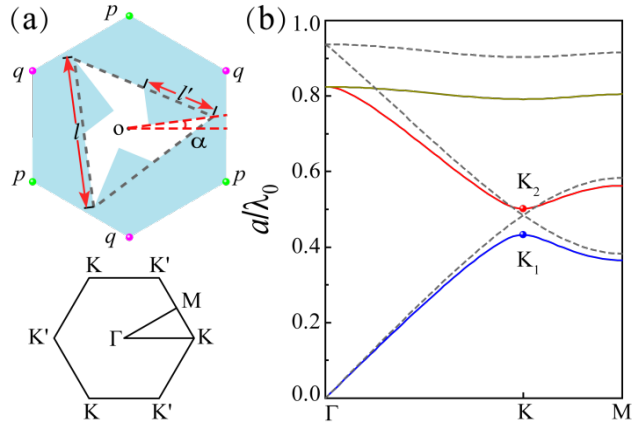

FIG. 1. (a) Unit cell diagram of a hexagonal SC (upper panel) and its first BZ (lower panel). The scatterer (white region) has a fan-like shape, which is tailored from a regular triangular scatterer (see the grey dashed frame). The angle $\alpha$ with respect to the horizontal axis characterizes the orientation of the scatterer. The points $p$ and $q$ label two inequivalent lattice centers with $C_3$ symmetry. (b) The band structure for the SC made of fan-like scatterers with $l'/l = 0.5$ (color solid lines), compared with that for the SC made of triangular scatterers (grey dashed lines). Here $a/\lambda_0$ depicts the dimensionless frequency, with $\lambda_0$ being the wavelength in air.

As shown in Fig. 1(a), our study starts from the SC system proposed previously [7], which consists of a hexagonal array of regular triangular scatterers (indicated by grey dashed lines). Specifically, the side length of the triangle $l = 0.78a$, with $a$ being the lattice constant. The orientation of the scatterer is characterized by the angle $\alpha$ with respect to the horizontal axis. It is easy to see that, if $\alpha = 0$ the SC has $C_{3v}$ symmetry, including a threefold rotational symmetry and three mirrors. The system supports deterministic Dirac cones at the inequivalent hexagonal corners of the first BZ, as shown by the grey dashed lines in Fig. 2(b). This stable double degeneracy is protected by the point group $C_{3v}$ that supports a 2D irreducible representation [7,19-21]. Now we consider the system with a lower symmetry, $C_3$, as shown in Fig. 1(a), where the triangular scatterer is modified into the fan-like one (white region) by removing three identical right triangles. The modulation strength can be characterized by a dimensionless geometry parameter $l'/l$, with $l'$ being the length of the hypotenuse of the right triangle. The color solid lines in Fig. 1(b) show the numerical dispersion curves for the fan-like system with a specific modulation strength $l'/l = 0.5$. As anticipated, the conic degeneracy is lifted since the fan-like system has only $C_3$ symmetry (and thus the point group supports only one-dimensional irreducible representations). The lowest two states at $K$ point correspond to a pair of frequency extrema in momentum space, i.e., the valley states $K_1$ and $K_2$, respectively, associated with an omnidirectional bandgap from the dimensionless frequency $0.43$ to $0.50$.

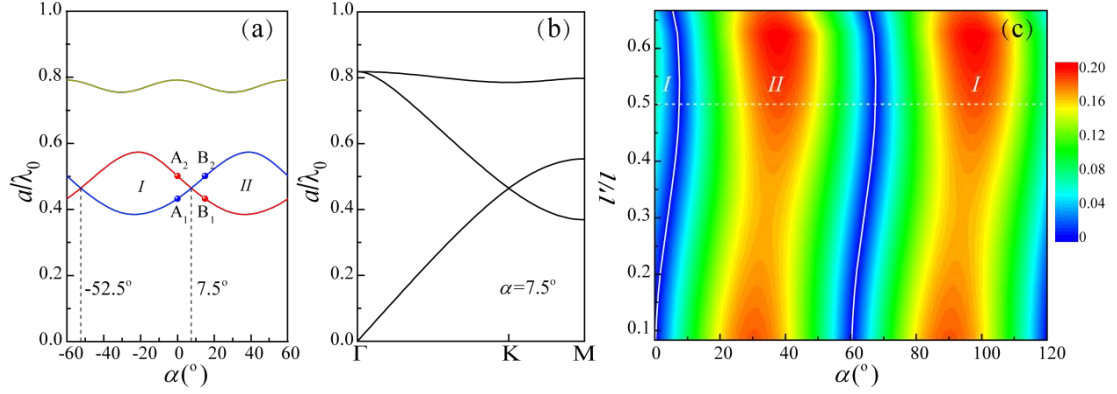

FIG. 2. (a) The band-edge frequencies depicted for the SCs with different $\alpha$ values within an angular period of $120°$. The angular regions *I* and *II* characterize different acoustic valley Hall phases concerned below. (b) The band structure for the SC with $\alpha = 7.5°$, indicating a linear cone at the $K$ point. (c) The frequency width of the first bandgap (color) plotted for the SCs made of fan-like scatterers with various angles and modulation strengths. The white solid line indicates the transition angle associated with zero bandgap, and the white dashed line marks the modulation strength $l'/l=0.5$ used before.

Double Dirac degeneracy may occur occasionally in the SC made of fan-like scatterers at some specific rotation angles. Figure 2(a) shows a continuous evolution of the band-edge frequencies (at $K$ point) versus the rotation angle $\alpha$. (Here only a reduced angular region of $2\pi/3$ is considered because of the $C_3$ symmetry of the fan-like scatterer.) It is observed that, the band-edge frequencies change smoothly, since the rotation tailors the scattering strength gradually among the scatterers. Interestingly, the first two bands cross with each other at the specific angles $\alpha \approx 7.5°$ and $\alpha \approx -52.5°$. At these angles, the band structures form double Dirac degeneracy at the hexagonal BZ corners. This is exemplified by $\alpha = 7.5°$ in Fig. 2(b), where the acoustic valley states $K_1$ and $K_2$ touch at the dimensionless frequency of 0.465. Note that the linear dispersion near the degenerated $K$ (or $K'$) point is guaranteed automatically, as theoretically proved in Ref. [7]. Without data shown here, Dirac degeneracy may also occur at higher frequencies. Therefore, the requirement of $C_{3v}$ symmetry for generating Dirac degeneracy is relaxed by utilizing the orientation degree of freedom of the scatterers. In fact, the rotating-scatterer mechanism is robust to realize accidental Dirac degeneracy at the fixed hexagonal BZ corners, as long as the scatterer has $C_3$ symmetry. To confirm this point, in Fig. 2(c) we plot the angular dependent bandgap width (color) for a wide range of modulation strengths $l'/l$. As manifested by the solid white lines, which depict the transition angles associated with gap closure, Dirac degeneracy always occurs at specific angles.

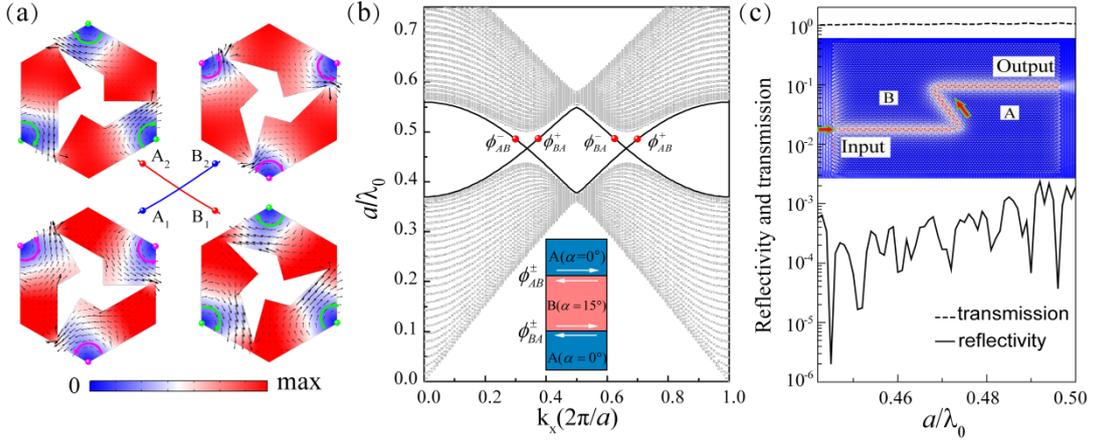

FIG. 3. (a) Eigen-field distributions of the first two valley states at $K$ point for the SCs with rotation angles $\alpha = 0^\circ$ (A, left) or $\alpha = 15^\circ$ (B, right), where the subscript 1 (2) denotes the first (second) band, as labeled in Fig. 2(a). The (green and magenta) dots and arrows indicate the distinct vortex centers and chiralities (i.e., orientations of the time-averaged Poynting vectors, see black arrows) of the acoustic valley states, respectively. (b) The interface dispersions (dark lines) evaluated for an A-B-A type superlattice structure, where the grey dots correspond to the bulk states. Inset: a schematic illustration for the interface structure, where the white arrows indicate the time-reversal pairs of edge states traveling along each interface. (c) The reflectivity and transmission spectra evaluated for the two sharp corners in a Z-shape interface channel. Inset: the sound profile evaluated at the dimensionless frequency 0.48.

It is of interest that the bandgap closure associated with accidental Dirac degeneracy suggests an acoustic valley Hall phase transition. Without losing generality, here we consider again the SCs with modulation strength $l'/l=0.5$, in which the transition angles $\alpha \approx 7.5^\circ$ and $\alpha \approx -52.5^\circ$ separate the angular regions $I$ and *II* [see Fig. 2(a)]. Figure 3(a) presents the eigenfield profiles of the acoustic valley states $K_1$ and $K_2$ evaluated for the SCs with $\alpha = 0^\circ$ and $\alpha = 15^\circ$ (briefly referred to as A and B, belonging to the angular regions *I* and *II*, respectively). All states display remarkable vortex features encircling the lattice centers $p$ or $q$, where the chiralities of the sound vortices are characterized by the orientations of the energy flows. Particularly, the order of the frequency bands (that carry given vortex features) is interchanged, a critical signature of the topological valley Hall phase transition. To confirm the acoustic valley Hall phase transition discuss above, we construct an A-B-A superlattice structure and calculate the projected band structures along the interfaces AB and BA simultaneously, as shown in Fig. 3(b). As expected, each interface hosts a time-reversal pair of edge states (dark lines), where $\phi_{AB}^\pm$ and $\phi_{BA}^\pm$ correspond to the edge states traveling on the interfaces AB and BA along the $\pm x$ directions. A fundamental property for the topologically nontrivial edge states is scattering immunity when suffering defects, which is much different from the situation emerging in a usual waveguide system. In Fig. 3(c) we demonstrate the negligibly weak backscattering of the acoustic valley Hall edge mode [21-23] propagating along a zigzag bending channel that separates the SCs A and B. As exemplified by the pressure distribution (see inset) simulated at the dimensionless frequency 0.48, the sound travels smoothly in the zigzag path despite suffering two sharp corners. In order to evaluate the reflection exactly from the two corners, a one-dimensional scattering model is employed to remove the multiple reflections from the exits [21]. As shown in Fig. 3(c), nearly full transmission occurs in the entire bulk gap region. It is worth noting that the rotating-scatterer mechanism enables easily tunable operation bandwidth and reconfigurable shape of the SC interface, which is very useful in real applications.

# III. Accidental quadruple Dirac degeneracy and acoustic quantum spin Hall phase transition

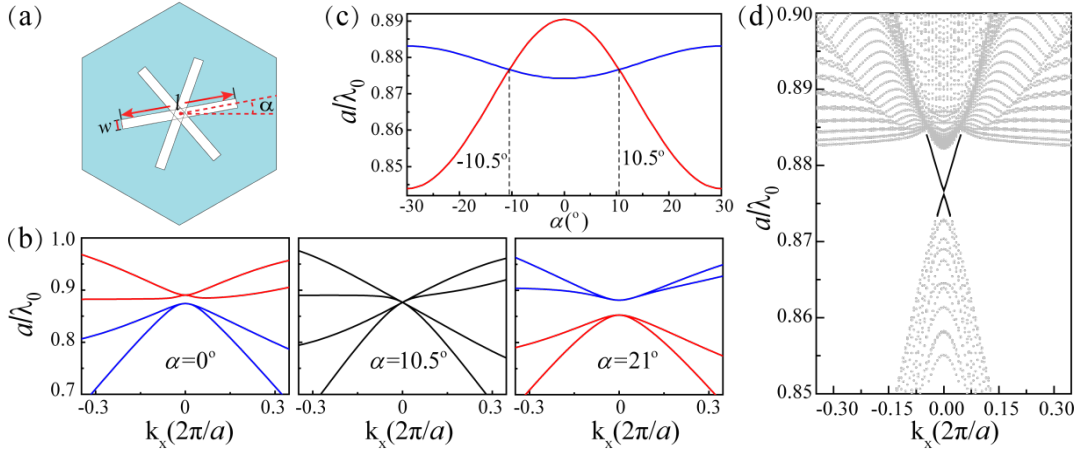

FIG. 4. (a) Unit-cell diagram of the SC consisting of a hexagonal array of snowflake-like scatterers. (b) Band structures (around the $\Gamma$ point) for three SCs with different rotation angles. (c) Doubly degenerated band-edge frequencies (at the $\Gamma$ point) depicted for the SCs with different $\alpha$ values. The result repeats at a period of $60^{\circ}$ considering the six-fold rotation symmetry of the system. (d) The projected band structures along an interface separating two SCs with $\alpha = 0^{\circ}$ and $\alpha = 21^{\circ}$.

The rotating-scatterer mechanism can also be used to produce the quadruple Dirac degeneracy at the center of the hexagonal BZ [12-15]. Recently, such a fourfold degeneracy, which links to a classical analog of the quantum spin Hall phase transition, has been realized by fine-tuning parameters [15,17,18] or folding BZ [24-26]. We consider a SC consisting of a hexagonal array of snowflake-like scatterers. Specifically, the length and width of the arms are $l = 0.6a$ and $w = 0.05a$, respectively. As shown in Fig. 4(a), the SC has $C_{6v}$ symmetry at the specific angles $\alpha = n\pi/3$, while the symmetry is reduced to $C_6$ in a general angle. Here we focus on the $\Gamma$ point of the hexagonal BZ, where quadratic (instead of linear) double degeneracy is generally guaranteed by the $C_6$ symmetry [7,24]. This can be exemplified by $\alpha = 0^{\circ}$ and $\alpha = 21^{\circ}$ [see Fig. 4(b)], where two quadratic double degeneracies emerge at the low frequency region, separating with an omnidirectional bandgap (associated with minimal gap at the $\Gamma$ point). Again, the bandgap can be tailored by rotating the scatterers. As shown in Fig. 4(c), the (doubly degenerated) band-edge frequencies vary smoothly and cross with each other at some specific angles (i.e., $\alpha \approx \pm 10.5^\circ$ at the reduced angular region). Once the crossing occurs, a quadruple Dirac degeneracy is generated at the hexagonal BZ center, as shown by $\alpha = 10.5^\circ$ in Fig. 4(b). Again, the gap closure and reopening correspond to a topological phase transition, similar to those observed previously [15,17,18,24-26]. To justify this point, we calculate the band structure for an interface separating two topologically distinct SCs, made of scatterers with $\alpha = 0^\circ$ and $\alpha = 21^\circ$. As shown in Fig. 4(d), interface dispersions appear in the bulk gap as predicted.

## IV. Conclusion

We have proposed a flexible rotating-scatterer mechanism to generate accidental double Dirac degeneracy at the corners of the hexagonal BZ. The SC associated with gap closure corresponds to a critical configuration of acoustic valley Hall phase transition. This is confirmed by the further observation of the gapless edge states along the interfaces separating topologically distinct acoustic valley Hall insulators. This approach has been further used to generate a quadruple Dirac degeneracy at the center of the hexagonal BZ. This gives a convenient realization for the classical analog of quantum spin Hall phase transition and the associated edge states [15,17,18,24-26].

**Acknowledgements**

This work is supported by the National Basic Research Program of China (Grant No. 2015CB755500); National Natural Science Foundation of China (Grant Nos. 11674250, 11534013, and 11547310); Postdoctoral innovation talent support program (BX201600054).